\documentstyle{article}
\begin{document}

\title{Quantum Theory Cannot Forbid Superluminal Signaling}
\author{{Li YANG}\\
   {\small State Key Laboratory of Information Security,}\\
   {\small Graduate School of
   Chinese Academy of Sciences,}\\
   {\small Beijing 100039,P. R. of China }}
\date{April 7,2001}
\maketitle

\begin{minipage}{110mm}
\vskip 0.3in

      This note analyzes the angular distributions of the probabilities of two-photon states
    come out of the single-photon's stimulated emission amplification by means of a single-atom
    amplifier, to see that whether the quantum theory can forbid us exploiting EPR photon pairs
    combined with stimulated emission to realize superluminal
    signaling. Besides, we stressed that superluminal signaling will leads to a dilemma of causality in a
    system with two long superluminal channels and two short
    light channels.
\end{minipage}
\vskip 0.5in

    It has been believed that the mathematical inseparability of
  the quantum theoretical representation is an essential part of
  nature, not a mere accident of the formalism[1]. However, the
  attempts to realize superluminal signaling over the last twenty
  years by means of EPR pairs has not been successful because one
  cannot clone a single particle in an unknown state sufficiently
  well [2]. Now let us see whether quantum theory can forbid
  superluminal signaling through a careful analysis of a concrete
  physical process.\\

    Consider the stimulated emission of a single excited atom, or a
  single-atom light amplifier[3]. Suppose the angle between the
  polarization direction of an incoming single-photon flow and the
  atom's transition dipole moment $\overrightarrow{\mu} $ is $\theta$   ,
   where $\overrightarrow{\mu} $  is perpendicular
  to the photon's wave vector. We note the photon's state by $|\theta\rangle$   .
  After scattering, the system's two-photon state is
  of the form[3]

       $$|\Psi^{\theta}_{f}\rangle=\alpha_{\theta}|2,0\rangle^{\theta}|g_{\theta}\rangle+
       \beta_{\theta}|1,1\rangle^{\theta}|g_{\theta+\frac{\pi}{2}}\rangle,\hskip
       1.5in
       (1)$$\\
  where $|2,0\rangle^{\theta}$    indicates that both photons are in the state $|\theta\rangle $  ,
  $|1,1\rangle^{\theta}$indicates a photon in each of the states $|\theta\rangle$
   and $|\theta+\frac{\pi}{2}\rangle$   , and $|g_{\theta}\rangle$
   and $|g_{\theta+\frac{\pi}{2}}\rangle$
  are the atom's final states. The state $|0,2\rangle^{\theta}$    would indicate that
  both photons are in the state $|\theta+\frac{\pi}{2}\rangle$   . Let $|\Omega\rangle^{\theta}$   be a vector with three
  components $|2,0\rangle^{\theta}$    ,$|0,2\rangle^{\theta}$   and $|1,1\rangle^{\theta}$   .
   We have $|\Omega\rangle^{\theta}=U(\theta)|\Omega\rangle^{0}$          , where $U(\theta)$   is a
  unitary transformation,
   $$|2,0\rangle^{\theta}=\cos^{2}\theta|2,0\rangle+\sin^{2}\theta|0,2\rangle+\frac{1}{\sqrt{2}}\sin2\theta|1,1\rangle,\hskip 0.3in(2a)$$

       $$|0,2\rangle^{\theta}=\sin^{2}\theta|2,0\rangle+\cos^{2}\theta|0,2\rangle-\frac{1}{\sqrt{2}}\sin2\theta|1,1\rangle,\hskip 0.2in(2b)$$

          $$|1,1\rangle^{\theta}=-\frac{1}{\sqrt{2}}\sin2\theta|2,0\rangle+\frac{1}{\sqrt{2}}\sin2\theta|0,2\rangle
                              +\cos2\theta|1,1\rangle.\hskip 0.2in(2c)$$

  Suppose $|g_{\theta}\rangle\neq|g_{\theta+\frac{\pi}{2}}\rangle$      and the
  incoming photons flow are a mixture of photons in $|\theta\rangle $ state and in $|\theta+\frac{\pi}{2}\rangle $
  state with the same probability. The four transition probabilities is
        $$w^{\theta}_{2,0}=2\lambda^{2}\cos^{2}\theta d\Omega,\hskip 1.0in(3a)$$
        $$w^{\theta}_{1,1}=\lambda^{2}\sin^{2}\theta d\Omega,\hskip 1.0in (3b)$$
        $$w^{\theta+\frac{\pi}{2}}_{0,2}=2\lambda^{2}\sin^{2}\theta d\Omega,\hskip 1.0in(3c)$$
        $$w^{\theta+\frac{\pi}{2}}_{1,1}=\lambda^{2}\cos^{2}\theta d\Omega.\hskip 1.0in (3d)$$
        where $w^{\theta}_{2,0}$ is the probability of
        generating a two-photon state $|2,0\rangle^{\theta}$ when incoming single-photon
        is in the state $|\theta\rangle$, $w^{\theta}_{1,1}$ is
        the probability of generating a two-photon state
        $|1,1\rangle^{\theta}$ when incoming single-photon is in
        the state $|\theta\rangle$, and so on; and
         $$\lambda^{2}=\frac{\omega^{3}\mu^{2}}{8\pi^{2}\hbar c^{3}}.\hskip 1.5in (4)$$
   Then by means of the unitary transformation $U(\theta)$ we can find the
  probability of the state $ |2,0\rangle  $ after scattering is
     $$d\overline{\sigma^{\theta}_{2,0}}=\frac{1}{2}\lambda^{2}(1+\cos^{2}2\theta)d\Omega.\hskip 1.0in(5)$$
  If we restrict our statistics to two-photon states, we have the
  probability of $|2,0\rangle$ state
     $$\overline{P^{\theta}_{2,0}}=\frac{1}{3}(1+\cos^{2}2\theta),\hskip 1.0in(6a)$$
  the other two probabilities are

     $$\overline{P^{\theta}_{1,1}}=\frac{1}{3},\hskip 1.4in(6b)$$
      $$ \overline{P^{\theta}_{0,2}}=\frac{1}{3}\sin^{2}2\theta.\hskip 1.0in(6c)$$

     [If let $|g_{\theta}\rangle=|g_{\theta+\frac{\pi}{2}}\rangle$, we
     have
        $$\overline{P^{\theta}_{2,0}}=\frac{2}{3}\cos^{2}2\theta,\hskip 1.0in(7a)$$
         $$\overline{P^{\theta}_{1,1}}=\frac{1}{3},\hskip 1.4in(7b)$$
         $$ \overline{P^{\theta}_{0,2}}=\frac{2}{3}\sin^{2}2\theta.\hskip
         1.0in(7c)]$$

       We can find from the analysis that one origin
       of the phenomenon that $\overline{P^{\theta}}$ is dependent on parameter $\theta$
       is the zero point energy of the light field, and the other one is likely to be
       the Bose-Einstein statistics of photons.\\

       The fact that $\overline{P^{\theta}}$ is dependent on the parameter $\theta$
       means that with the single-atom light
     amplifier one can distinguish the parameter $ \theta$ of an
     incoming single-photon flow by measuring the probabilities of
     $|2,0\rangle$ state or $|0,2\rangle$ state. If Alice and Bob share a sufficiently
     large number of two-photon EPR pairs in the Bell states with
     rotation invariance, they do not need another channel to
     complete their communication, then there is no law of physics
     which will obviously stop the superluminal signaling between
     Alice and Bob.\\

       It can be shown that the superluminal signaling between long distance leads to
     a dilemma of causality in a system with two pairs of Alice and Bob belonged to
      two different inertial frames separately, and with four channels: two long
      superluminal channels [Alice(1),Bob(1)] and [Alice(2),Bob(2)],and two short
      light channels [Bob(1),Alice(2)]and [Bob(2),Alice[1]]. Then we know that
      if any phenomenon of superluminal signaling be found in experiment or theoretical analysis,
      it means directly something wrong with our understanding of nature.\\

             Because of the success of Electrodynamics, General
           Relativity, and Quantum Field Theories we believe that
           the Lorentz covariance of our theories is right. Based
           on Lorentz covariance we can prove that any
           superluminal signaling will lead to the dilemma of
           causality. Then we believe that there is no
           superluminal signaling in nature. However, this
           conclusion does not imply that any non-relativistic
           theory cannot give out a result related with
           superluminal signaling at all. It is difficult to
           understand some people's belief that the
           non-relativistic quantum theory itself can forbid
           superluminal signaling automatically. One may argue
           that he believes there is no conflict between the
           theory of relativity and the theory of quantum
           mechanics, but many people(for example, J. A. Wheeler) believe that the conflict is
           unavoidable. Considering the great difficulties we meet in the
           quantum field theory, especially in the quantum theory of
           gravity, we should realize that the conflict is evident
           and essential.

\end{document}